\documentclass{IEEEtran}
\usepackage{cite}
\usepackage{amsmath,amssymb,amsfonts}
\usepackage{algorithmic}
\usepackage[dvipsnames]{xcolor}
\usepackage{graphicx}
\usepackage{subcaption}
\usepackage{textcomp}
\def\BibTeX{{\rm B\kern-.05em{\sc i\kern-.025em b}\kern-.08em
    T\kern-.1667em\lower.7ex\hbox{E}\kern-.125emX}}

\newtheorem{remark}{Remark}[section]

\begin{document}
\title{Physics--guided gated recurrent units for inversion--based feedforward control}
\author{Mingdao Lin, Max Bolderman, \IEEEmembership{Member, IEEE}, and Mircea Lazar, \IEEEmembership{Senior Member, IEEE}
\thanks{This work was supported by the NWO, The Netherlands, research project PGN Mechatronics, project number 17973.}
\thanks{M. Lin, M. Bolderman and M. Lazar were affiliated with the Eindhovern University of Technology, Control Systems Group during the period when this research was conducted (e--mails: lmdholland@outlook.com, max.bolderman@hotmail.com, m.lazar@tue.nl). }}

\maketitle
\begin{abstract}
Inversion--based feedforward control relies on an accurate model that describes the inverse system dynamics. 
The gated recurrent unit (GRU), which is a recent architecture in recurrent neural networks, is a strong candidate for obtaining such a model from data.
However, due to their black--box nature, GRUs face challenges such as limited interpretability and vulnerability to overfitting. 
Recently, physics--guided neural networks (PGNNs) have been introduced, which integrate the prior physical model structure into the prediction process. 
This approach not only improves training convergence, but also facilitates the learning of a physics--based model. 
In this work, we integrate a GRU in the PGNN framework to obtain a PG--GRU, based on which we adopt a two--step approach to feedforward control design.
First, we adopt stable inversion techniques to design a stable linear model of the inverse dynamics. 
Then, a GRU trained on the residual is tailored to inverse system identification. 
The resulting PG--GRU feedforward controller is validated by means of real--life experiments on a two--mass spring--damper system, where it demonstrates roughly a two--fold improvement compared to the linear feedforward and a preview--based GRU feedforward in terms of the integral absolute error. 
\end{abstract}

\begin{IEEEkeywords}
Feedforward control, gated recurrent units, motion control, recurrent neural networks.
\end{IEEEkeywords}

\section{Introduction}
In high--precision motion control, improving both \emph{accuracy} and \emph{throughput} remains a key objective. 
Typically, a two--degree--of--freedom (2--DoF) control structure is adopted, in which feedback control ensures closed--loop stability and disturbance rejection~\cite{iwasaki2012high, boerlage2003model}. 
In addition, feedforward control achieves high reference tracking performance by compensating for reference before an error occurs~\cite{van2018inversion}. 

Inversion--based feedforward control designs an input by passing the reference through a known model that describes the inverse system dynamics.
When the system is non--minimum phase, such an inverse model becomes unstable rendering the feedforward controller not useful in practice. 
To address this problem, stable inversion techniques have been designed, including approximation such as non--minimum--phase zeros ignore (NPZ--Ignore), zero--magnitude--error tracking controller (ZMETC), or zero--phase--error tracking controller (ZPETC)~\cite{butterworth2012analysis}. 
Alternative to approximation, it is possible to design a non--causal feedforward controller when the complete reference is known a priori. 
These techniques apply to linear models, while real--life systems exhibit parasitic effects.
Therefore, using linear models for feedforward control implicitly limits the achievable performance. 

With the aim to increase the accuracy, nonlinear models such as neural networks (NNs) are used to approximate the inverse system dynamics.
For example, nonlinear autoregressive networks with exogenous inputs (NNARXs) have been proposed in~\cite{lin1996learning, sorensen1999additive, bolderman2021physics}, which use an input--output model.
However, not all systems admit an input--output representation~\cite{pearson2004nonlinear}, which again induces structural modeling errors. 
This limitation has led to the development of state--space neural networks (SSNNs) for system identification \cite{schoukens2021improved, wang2006fully,forgione2023adaptation}. 
Traditional recurrent neural networks (RNNs) utilize hidden states to track historical information but struggle to capture long--term dependencies due to the vanishing gradient problem \cite{bengio1994learning}. 
Long short--term memory networks (LSTMs) address this issue with a three--gate architecture but incur a four--fold parameter increase \cite{hochreiter1997long}, \cite{jozefowicz2015empirical}. 
Gated recurrent units (GRUs) offer a more efficient alternative with only two gates \cite{chung2014empirical}, which have been used as an approximate feedforward compensation term in~\cite{hu2020gru, hu2020gruFF,zhou2023intelligent}, by adding the predicted tracking error to the reference. 
In model predictive control, GRUs perform comparably to LSTMs but with fewer parameters \cite{bonassi2021nonlinear,zarzycki2022advanced,Bonassi2024GRUMPC}. 
While Transformer architectures are another option, their high computational cost and inference latency make them impractical for real--time control.
Moreover, stability proofs exist for GRUs and LSTMs \cite{bonassi2021stability,bonassi2020lstm}, but currently not for Transformers.

Despite the potentially improved feedforward control performance, both input--output NNs and SSNNs share common drawbacks: limited model interpretability and non--robust training processes. 
To address these drawbacks and enhance compliance with physical principles, physics--informed neural networks (PINNs) \cite{karpatne2017physics} and physics--guided neural networks (PGNNs) \cite{bolderman2021physics,bolderman2024physics} have been developed. 
A PGNN integrates a known physical model with an NN in a single model to predict the output, while PINNs incorporate physical laws into the cost function. 
Both approaches improve training convergence and help to capture the underlying physics.
However, their standard form adopts NARX--type formulations without explicit state representations.
Thus, they cannot fully capture nonlinear state--space dynamics when applied to systems whose behavior depends on latent state evolution.

To solve the aforementioned limitations, in this work we develop a PG--GRU feedforward controller, which combines a linear model and a GRU.
Thereby, the resulting contributions of this work are summarized as follows:
\begin{enumerate}
    \item We adopt a preview window in the GRU model for identification of the inverse system; 
    \item We combine a (stabilized) linear feedforward controller with the GRU to obtain a physics--guided GRU (PG--GRU) feedforward controller;
    \item We validate the PG--GRU on a non--minimum phase two--mass spring--damper experimental setup.
\end{enumerate}

The remainder of this work is organized as follows: Section~\ref{section Preliminaries and Problem statement} introduces the control scheme, the feedforward control design, and the problem statement. 
The PG--GRU feedforward control design is presented in Section~\ref{sec:PGGRU_Feedforward}, followed by the experimental results in Section~\ref{section results}. 
Finally, the main conclusions and future research directions are summarized in Section~\ref{section conclusions and discussions}. 

\section{Preliminaries and Problem statement}
\label{section Preliminaries and Problem statement}
\subsection{Feedback-feedforward control architecture}
\begin{figure}
    \centering \includegraphics[width=1.0\linewidth]{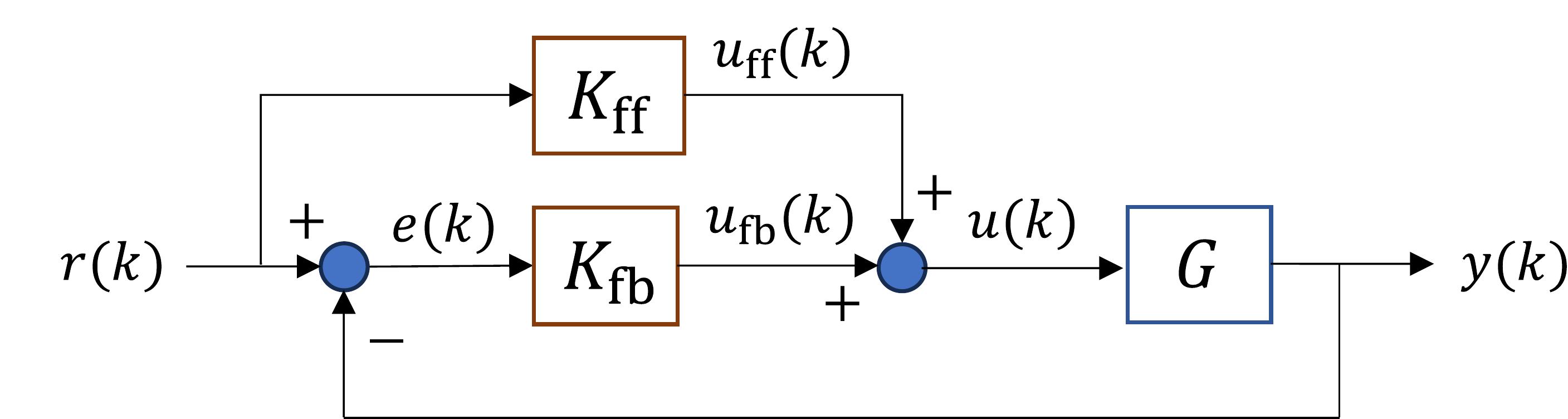}
    \caption{A standard discrete--time 2--DoF control architecture.}\label{figure: 2dof_DT}
\end{figure}
Fig.~\ref{figure: 2dof_DT} shows the two--degree--of--freedom control structure, with $K_{\textup{fb}}$, $K_{\textup{ff}}$ and $G$ the feedback controller, the feedforward controller and the system.
The discrete--time instant is represented by $k \in \mathbb{Z}_{>0}$. 
The input is denoted by $u(k) \in \mathbb{R}^{n_u}$, the output is $y(k) \in \mathbb{R}^{n_y}$, and the reference is $r(k) \in \mathbb{R}^{n_y}$, with $n_u, n_y\in \mathbb{Z}_{>0}$. The tracking error is defined as $e(k):=r(k)-y(k)$. 
The state of the system is represented by \(x(k) \in \mathbb{R}^{n_x}\), with \(n_x \in \mathbb{Z}_{>0}\). 
Consider a strictly proper, linear discrete--time multi--input multi--output (MIMO) system.
Then the closed--loop dynamics is
\begin{align}
    \begin{split}
        \label{eq:ClosedLoop_System}
        x(k+1) & = A x(k) + B u(k), \\
        y(k) & = C x(k), \\
        u(k) & = K_{\textup{fb}}(z) \big( r(k)-y(k) \big) + u_{\textup{ff}}(k), 
    \end{split}
\end{align}
with $A\in\mathbb R^{n_x\times n_x}$, $B\in\mathbb R^{n_x\times n_u}$, and $C\in\mathbb R^{n_y\times n_x}$. $u_{\textup{ff}}(k) \in \mathbb{R}^{n_u}$ is the feedforward input, and the feedback input is given as $u_{\textup{fb}}(k) = K_{\textup{fb}} (z) e(k) $, with $z$ the forward shift operator, e.g., $e(k) = z\cdot e(k-1) = z^2\cdot e(k-2)$ and $K_{\textup{fb}}(z)$ a rational transfer function \cite{bolderman2024physics}. 
Next, we assume that perfect tracking is achieved such that $e(k) = r(k)-y(k) = \mathbf{0}\in\mathbb R^{n_y}, \forall k \in \mathbb{Z}_{>0}$, which gives $u_{\textup{fb}}(k) = K_{\textup{fb}}(z) e(k) = \mathbf{0}\in\mathbb R^{n_u}$. 
As a result, the feedforward controller satisfies:
\begin{align}
    \begin{split}
        \label{eq:Feedforward_Step1}
        x_{\textup{ff}}(k+1) & = A x_{\textup{ff}}(k) + B u_{\textup{ff}}(k) , \\
        r(k) & = C x_{\textup{ff}}(k),
    \end{split}
\end{align}Let $\eta_0 \in \mathbb{Z}_{> 0}$ be the relative degree, i.e., the smallest value for which $CA^{\eta_0-1}B$ is contains a non--zero entry, and assume that $CA^{\eta_0-1}B$ is invertible/non--singular.
Then, from~\eqref{eq:Feedforward_Step1} we have $r(k+\eta_0) = A^{\eta_0} x_{\textup{ff}} (k) + C A^{\eta_0-1} B u_{\textup{ff}}(k)$, such that we obtain the feedforward controller
\begin{align}
\begin{split}
    \label{eq:FeedforwardController}
        x_{\textup{ff}}(k+1) & = A_{\textup{ff}} x_{\textup{ff}}(k) + B_{\textup{ff}} r(k+\eta_0), \\
        u_{\textup{ff}}(k) & = C_{\textup{ff}} x_{\textup{ff}}(k) + D_{\textup{ff}} r(k+\eta_0),
\end{split}
\end{align}
where
\begin{align}
    \begin{split}
        \label{eq:Feedforward_Matrices}
        A_{\textup{ff}} & = A - B(CA^{\eta_0-1}B)^{-1} C A ^{\eta_0}, \\
        B_{\textup{ff}} & = B(CA^{\eta_0-1}B)^{-1} , \\
        C_{\textup{ff}} & = -(CA^{\eta_0-1}B)^{-1} C A ^{\eta_0} , \\
        D_{\textup{ff}} & = (CA^{\eta_0-1}B)^{-1}. 
    \end{split}
\end{align}
The feedforward controller~\eqref{eq:FeedforwardController},~\eqref{eq:Feedforward_Matrices} can be implemented directly when the system matrices $A$, $B$ and $C$ are known, and:
\begin{enumerate}
    \item \emph{Preview:} the reference $r(k+\eta_0)$ is known at time $k$;
    \item \emph{Stability:} the eigenvalues of $A_{\textup{ff}}$ are within the unit circle. 
\end{enumerate}
When the matrix $A_{\textup{ff}}$ has eigenvalues outside the unit circle, it is common practice to approximate the unstable poles or to perform a non--causal inversion. 
To illustrate these approaches, we emphasize that we consider a single--input single--output (SISO) case for ease of demonstration, and rewrite the state--space feedforward controller~\eqref{eq:FeedforwardController} in transfer function notation:
\begin{align}
    \begin{split}
        \label{eq:FeedforwardController_TransferFunction2}
        u_{\textup{ff}}(k) & = \Big( C_{\textup{ff}} (zI-A_{\textup{ff}})^{-1} B_{\textup{ff}} + D_{\textup{ff}} \Big)r(k+\eta_0)\\
        & = \frac{1}{\Pi_{i=1}^{n_{\textup{us}}} (z-p_i)} \tilde{K}_{\textup{ff}}(z) r(k+\eta_0),
    \end{split}
\end{align}
where $\tilde{K}_{\textup{ff}}(z)$ is the stable part of the feedforward controller and $p_i$ are the unstable poles, $i = 1, ..., n_{\textup{us}}$ and $n_{\textup{us}} \in \mathbb{Z}_{> 0}$ the number of unstable poles.
$p_i$ is either non--causally computed, or approximated using, e.g., ZPETC~\cite{van2018inversion}.
This yields:
\begin{align}
    \begin{split}
        \label{eq:StableInversion}
        \textup{Non--causal:} \quad\; & \frac{1}{z-p_i} \approx -\frac{1}{p_i} - \frac{1}{p_i^2} z - \frac{1}{p_i^3} z^2 - ... \; , \\
        \textup{ZPETC:} \quad \; & \frac{1}{z-p_i} \approx \frac{z^{-1} -p_i}{(1-p_i)^2} = \frac{1 -p_i z}{(1-p_i)^2z } . 
    \end{split}
\end{align}
ZPETC approximation requires an additional preview sample for every unstable pole in the feedforward controller~\eqref{eq:FeedforwardController_TransferFunction2}, and the non--causal design requires an preview of the complete reference profile.
Nevertheless, the non--causal design can be truncated as the coefficients become smaller (note, $|p_i| > 1$ for it to be unstable). 
After stable approximation of the unstable poles, the feedforward controller is given as
\begin{equation}
    \label{eq:FeedforwardController_TransferFunction}
    u_{\textup{ff}}(k) = K_{\textup{ff}}(z) r(k+\eta_0+n_{\textup{ep}} ) ,
\end{equation}
where $n_{\textup{ep}} \in \mathbb{Z}_{\geq 0}$ is the number of extended preview samples induced by the stable inversion.

\subsection{Identification for feedforward control}
Based on first-principle modeling, a parametrized model of  system~\eqref{eq:ClosedLoop_System} is constructed as
\begin{align}
    \begin{split}
        \label{eq:ClosedLoop_Model}
        \hat{x}(k+1) & = A (\theta_{\textup{phy}}) \hat{x}(k) + B (\theta_{\textup{phy}}) \hat{u}(k), \\
        \hat{y}(k) & = C \hat{x}(k), \\
        \hat{u}(k) & = K_{\textup{fb}}(z) \big( r(k) - \hat{y}(k) \big) + u_{\textup{ff}}(k),
    \end{split}
\end{align}
where a hat indicates a model--based prediction, e.g., $\hat{y}(k)$ is a prediction of the output $y(k)$ at time~$k$, and $\theta_{\textup{phy}} \in \mathbb{R}^{n_{\theta_{\textup{phy}}}}$ denotes the parameters corresponding to physical quantities, such as inertia, damping, and stiffness coefficients.
These parameters are initialized via curve fitting of measured frequency response data \cite{kasemsinsup2018experimental}, then refined by optimizing:
\begin{equation}
    \label{eq:Identification_Criterion}
    \hat{\theta} = \textup{arg} \min_{\theta} \frac{1}{N} \sum_{k =1}^N \big( y^d(k) - \hat{y}^d(k) \big)^2, 
\end{equation}
where the superscript $d$ denotes that a variable corresponds to a data set. 
The data set is denoted as $Z^N$, with $N \in \mathbb{Z}_{>0}$ is the number of samples, and is given as
\begin{equation}
    \label{eq:DataSet}
    Z^N = \{ r^d(1), u_{\textup{ff}}^d(1), y^d(1), ..., r^d(N), u_{\textup{ff}}^d(N), y^d(N) \}. 
\end{equation}

It was shown in~\cite{bolderman2023data} that this closed--loop identification aims to find the parameters that yield the best tracking performance when using the identified model for feedforward control.  
The feedforward controller~\eqref{eq:FeedforwardController} designed from the identified model~\eqref{eq:ClosedLoop_Model} is obtained by using the matrices $A(\hat{\theta}_{\textup{phy}})$, $B(\hat{\theta}_{\textup{phy}})$, and $C(\hat{\theta}_{\textup{phy}})$ in~\eqref{eq:Feedforward_Matrices}. 

\begin{figure}
    \centering 
    \includegraphics[width=0.9\linewidth]{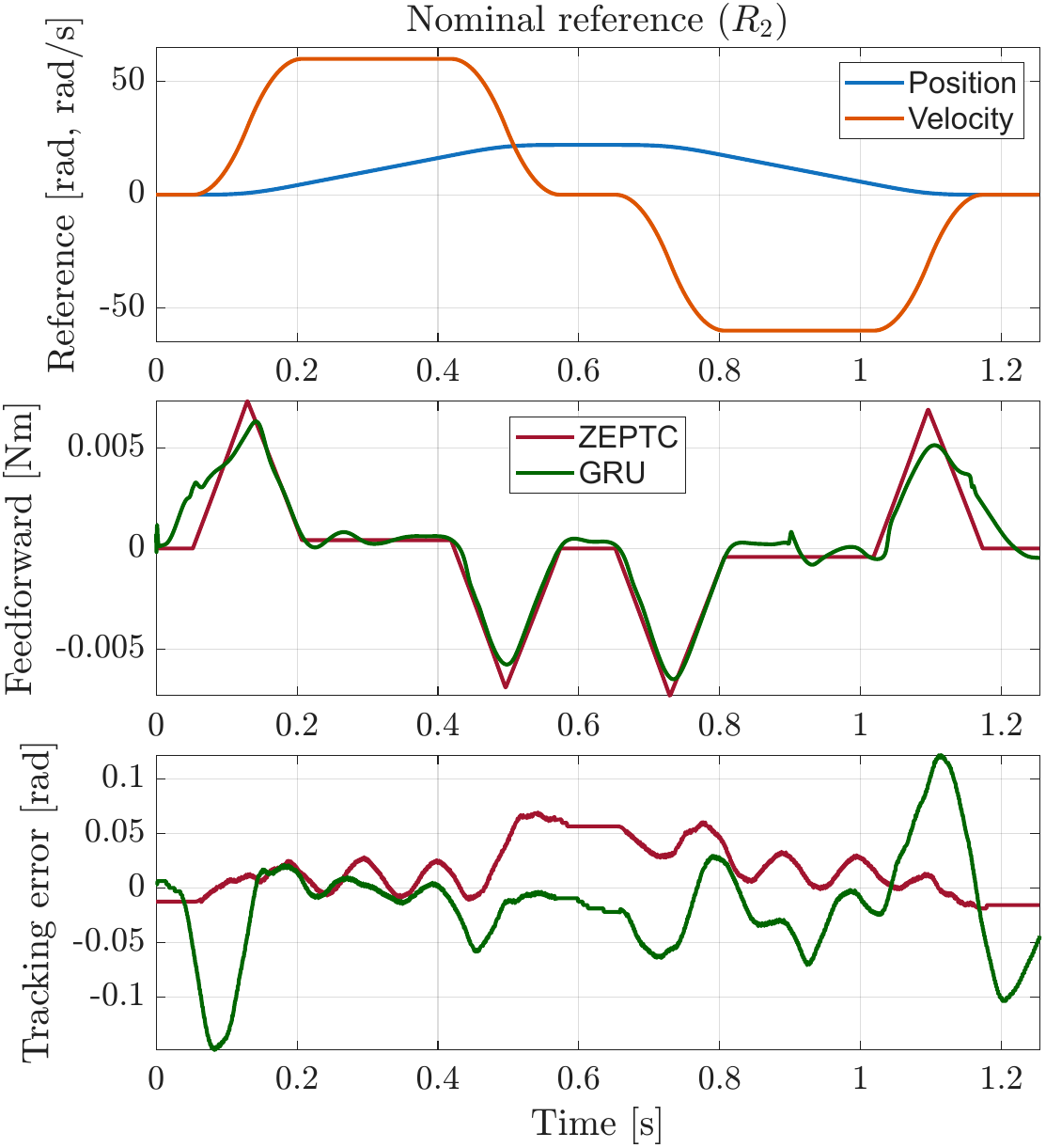}
    \caption{Feedforward control experiment on the two--mass spring--damper system, from top to bottom: reference, feedforward inputs, and tracking errors using ZPETC feedforward ({\color{Maroon}\textbf{{---}}}) and GRU feedforward ({\color{ForestGreen}\textbf{{---}}}). Details regarding the system and experiment are provided in Section~\ref{section results}.}  
    \label{fig:zpetc gru ff}
\end{figure}

\subsection{Problem statement}
Fig.~\ref{fig:zpetc gru ff} shows the tracking error for a closed--loop experiment on a two--mass spring--damper system that is further explained in Section~\ref{section results}. 
We adopt a linear feedforward controller~\eqref{eq:FeedforwardController_TransferFunction} with ZPETC stable inversion~\eqref{eq:StableInversion} and a standard GRU NN that is trained to replicate the inverse system as in~\cite{bolderman2021physics}. 
The linear ZPETC feedforward controller achieves limited accuracy, since real--life systems exhibit parasitic nonlinear effects that are not included in the model. 
In particular, these effects are often state--dependent, which motivates the exploration of feedforward control strategies using nonlinear models with internal states, such as SSNNs to further enhance performance.
However, from Fig.~\ref{fig:zpetc gru ff} we observe that using directly a black--box GRU does not improve performance as it fails to identify the main system dynamics. 

Following the aforementioned discussion, the objective of this work is to enhance the design and performance of inversion--based feedforward controllers by combining the linear feedforward control design with a GRU. 
To achieve this, we propose PG--GRUs, which incorporate a stable linear inverse model, and use a GRU tailored for inverse identification to approximate the residual. 
Since the GRU will follow an inverse identification, i.e., where the measured output becomes the input, we implement a filter to lower noise and quantization effects of the measurements. 
\begin{figure*}
    \centering \includegraphics[width=0.8\linewidth]{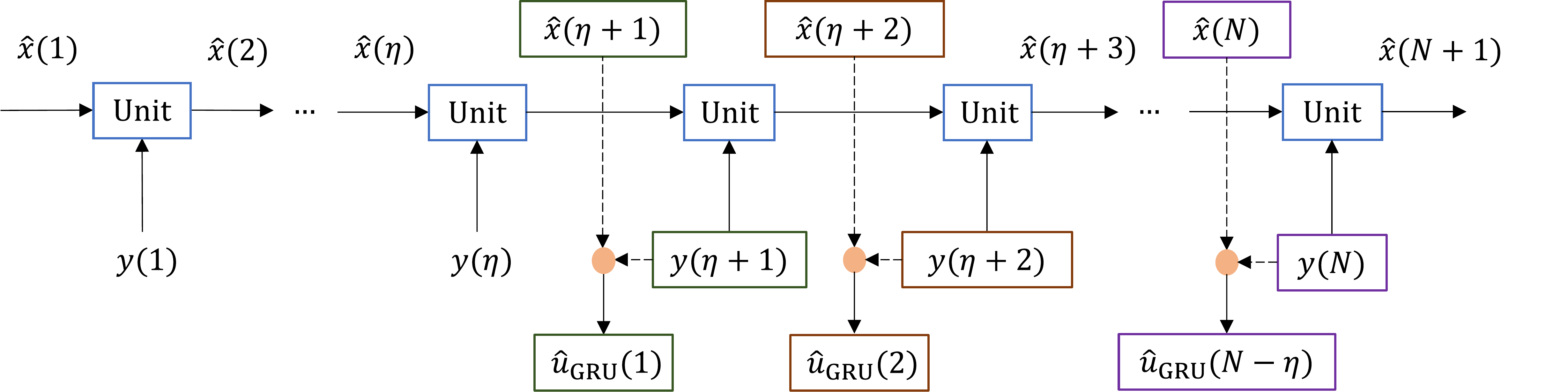}
    \caption{A schematic overview of the GRU model with preview in unfolded form. Above, the symbol {\color{Peach}\textbf{{$\bullet$}}} denotes a linear combination of the inputs, i.e., the output signals are generated by a weighted summation of input signals.}\label{fig:GRU_preview_unfolded}
\end{figure*}

\section{PG--GRU feedforward control design}
\label{sec:PGGRU_Feedforward}
\begin{figure*}
    \centering \includegraphics[width=0.65\linewidth]{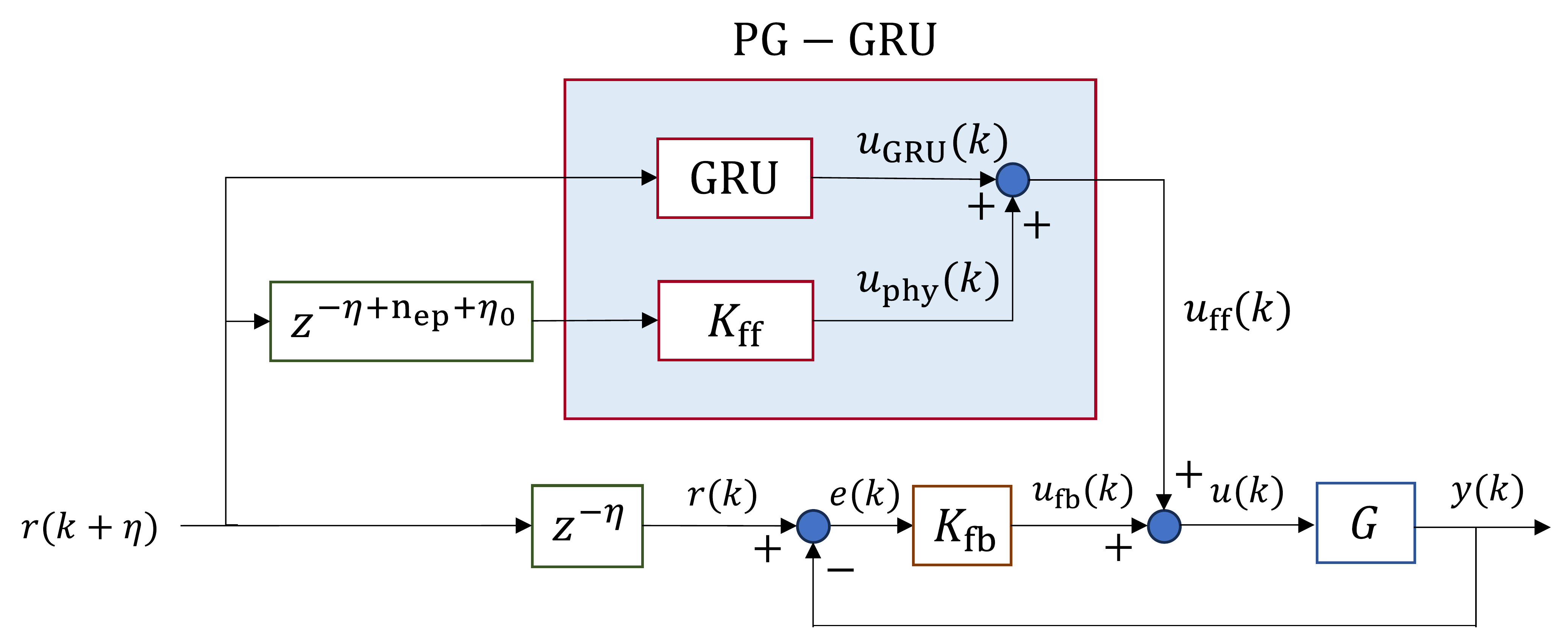}
    \caption{Schematic overview of the implementation of the PG--GRU feedforward~\eqref{eq:Feedforward_PGGRU}. Above, the symbol {\color{RoyalBlue}\textbf{{$\bullet$}}} denotes the direct summation of the inputs.}
    \label{ZPETC-GRU FF}
\end{figure*}
\subsection{Preview--based GRU for feedforward control}
\label{section tf gru FF}
Due to the strict--causality of the system~\eqref{eq:ClosedLoop_System}, the feedforward controller~\eqref{eq:FeedforwardController_TransferFunction} requires a preview of $\eta_0$ samples. 
Additionally, the stable inversion such as given in~\eqref{eq:StableInversion}, typically further extends the preview window for non--minimum phase systems. 
Let $\eta \geq \eta_0$ denote the preview window. Hence, the GRU model with preview of $\eta$ to model the inverse system dynamics as shown in Fig.~\ref{fig:GRU_preview_unfolded}, is formulated as:
\begin{align}
    \begin{split}
        \label{eq:GRU_Model}
        \hat{x}(k+1) & = \hat{z}(k) \circ \hat{x}(k) + \big( 1 - \hat{z}(k) \big) \circ \phi \big( W_x y(k) \\
        & \quad + U_x \hat{s}(k) \circ \hat{x}(k) + b_x \big), \\
        \hat{u}_{\textup{GRU}}(k) & = W_u y(k+\eta) + U_u \hat{x}(k + \eta) + b_u, \\
        \hat{z}(k) & = \sigma \big( W_z y(k) + U_z \hat{x}(k) + b_z \big), \\
        \hat{s}(k) & = \sigma \big( W_s y(k) + U_s \hat{x}(k) + b_s \big). 
    \end{split}
\end{align}
In~\eqref{eq:GRU_Model}, $\phi : \mathbb{R}^{n_{\textup{GRU}}} \rightarrow \mathbb{R}^{n_{\textup{GRU}}}$ represents the element--wise activation function, with $n_{\textup{GRU}}\in \mathbb{Z}_{>0}$ the number of neurons, $\circ$ is the Hadamard product, and $\hat{z}(k)$, $\hat{s}(k)$ the update and reset gates. 
The set of parameters of the GRU~\eqref{eq:GRU_Model} are
\begin{equation}
    \label{eq:GRU_Parameters}
    \theta_{\textup{GRU}} = \{ W_x, U_x, b_x, W_u, U_u, b_u, W_z, U_z, b_z, W_s, U_s, b_s \}. 
\end{equation}
\begin{remark}
Unlike standard types of GRU used in literature, see, e.g., \cite{chung2014empirical,hu2020gru,zhou2023intelligent, bonassi2021nonlinear,zarzycki2022advanced}, the GRU in~\eqref{eq:GRU_Model} computes $\hat u_\text{GRU}(k)$ as a function of $y(k+\eta)$ since it includes a preview window $\eta$. Here, $\eta$ is also considered a hyperparameter.
\end{remark}

\subsection{Training the preview--based GRU}
The preview--based GRU~\eqref{eq:GRU_Model} models the inverse of the open--loop system. 
To identify its parameter $\theta_{\textup{GRU}}$, an input--output data set generated on the system~\eqref{eq:ClosedLoop_System} is given as
\begin{equation}
    \label{eq:DataSet_InputOutput}
    Z^N = \{ u^d(1), y^d(1), ..., u^d(N), y^d(N) \}. 
\end{equation}
Note that, from~\eqref{eq:GRU_Model}, the GRU can predict up until $\hat{u}_{\textup{GRU}}^d(N-\eta)$ when given the data set $Z^N$. 
Moreover, the state $\hat{x}(k)$ of the GRU does not constitute any physical interpretation. 
Therefore, we do not know how to initialize~$\hat{x}(0)$. 
Although some approaches focus on parameterizing another neural network to predict $\hat{x}(0)$, see, e.g.,~\cite{beintema2021nonlinear}, we follow a more ad hoc approach.
Namely, since we are interested in a stable GRU model, mismatches in the initial state will converge to zero. 
For this reason, we exclude the first $\beta \in \mathbb{Z}_{>0}$ in the cost function.
Here, $\beta$ is considered a hyperparameter.
The resulting identification criterion is given as
\begin{align}
\begin{split}
    \label{eq:IdentificationCriterion_GRU}
    \hat{\theta}_{\textup{GRU}} = \textup{arg} \min_{\theta_{\textup{GRU}}} & \frac{1}{N - \eta - \beta} \sum_{k = 1 + \beta}^{N - \eta} \big( u^d(k) - \hat{u}_{\textup{GRU}}^d (k) \big)^2 \\
    & + \lambda \| \theta_{\textup{GRU}} \|_2^2. 
\end{split}
\end{align}
In~\eqref{eq:IdentificationCriterion_GRU}, $\|\cdot \|_2^2$ represents the squared $2$--norm, such that $\lambda \in \mathbb{R}_{\geq 0}$ represents the amount of $L2$ regularization. 

Finally, the GRU--based feedforward controller is obtained by substituting $\theta_{\textup{GRU}} = \hat{\theta}_{\textup{GRU}}$, $ y(k) = r(k)$, $\hat{x}(k) = x_{\textup{ff}}(k)$ and $\hat{u}_{\textup{GRU}} (k) = u_{\textup{ff}}(k)$ in~\eqref{eq:GRU_Model}, such that we obtain
\begin{align}
    \begin{split}
        \label{eq:GRU_Feedforward}
        x_{\textup{ff}} (k+1) & = z(k) \circ x_{\textup{ff}}(k) + \big( 1 - z(k) \big) \circ \phi \big( \hat{W}_x r(k) \\
        & \quad + \hat{U}_x s(k) \circ x_{\textup{ff}}(k) + \hat{b}_x \big) , \\
        u_{\textup{ff}}(k) & = \hat{W}_u r(k+\eta) + \hat{U}_x x_{\textup{ff}} (k + \eta) + \hat{b}_u, \\
        z(k) & = \sigma \big( \hat{W}_z r(k) + \hat{u}_z x_{\textup{ff}}(k) + \hat{b}_z \big) , \\
        s(k) & = \sigma \big( \hat{W}_s r(k) + \hat{U}_s x_{\textup{ff}}(k) + \hat{b}_s \big). 
    \end{split}
\end{align}

\subsection{Physics--guided GRU feedforward control}
\label{section pggru ff}
We adopt the GRU~\eqref{eq:GRU_Model} to learn only the inverse system dynamics that is not captured by the linear model~\eqref{eq:ClosedLoop_Model}.
Note that, the linear feedforward controller~\eqref{eq:FeedforwardController_TransferFunction} is a stable, linear model of the inverse system. 
By replacing $r(k) = y^d(k)$, we obtain the linear, physics--based prediction of the input as
\begin{equation}
    \label{eq:InverseModel_Linear}
    \hat{u}_{\textup{phy}}^d (k) = K_{\textup{ff}}(z) y^d(k+\eta_0 + n_{\textup{ep}}). 
\end{equation}
Next, we define the residual according to
\begin{equation}
    \label{eq:Residuals}
    \varepsilon^d (k) = u^d(k) - \hat{u}_{\textup{phy}}^d(k), \quad k = 1, ..., N -\eta_0 - n_{\textup{ep}}. 
\end{equation}
We train the GRU~\eqref{eq:GRU_Model} to learn the residual~\eqref{eq:Residuals}, such that
\begin{align}
\begin{split}
    \label{eq:IdentificationCriterion_PGGRU}
    \hat{\theta}_{\textup{GRU}} = \textup{arg} \min_{\theta_{\textup{GRU}}} & \frac{1}{N - \eta - \beta} \sum_{k = 1 + \beta}^{N - \eta} \big( \varepsilon^d(k) - \hat{u}_{\textup{GRU}}^d (k) \big)^2 \\
    & + \lambda \| \theta_{\textup{GRU}} \|_2^2. 
\end{split}
\end{align}
Finally, the feedforward controller is given as
\begin{equation}
    \label{eq:Feedforward_PGGRU}
    u_{\textup{ff}}(k) = u_{\textup{phy}}(k) + u_{\textup{GRU}}(k), 
\end{equation}
with $u_{\textup{phy}}(k)$ the feedforward input in~\eqref{eq:FeedforwardController_TransferFunction} and $u_{\textup{GRU}}(k)$ the feedforward input in~\eqref{eq:GRU_Feedforward}. 
\begin{remark}
    A key advantage of the PG--GRU is that the GRU is that the majority of the feedforward input results from the linear, physics--based model. 
    The GRU is used for the, relatively smaller, mismatches. 
    Thereby, the GRU contribution in the PG--GRU is significantly smaller compared to using a stand--alone GRU, which enhances interpretability. 
    Table~\ref{tab:pggru_traditional} summarizes the main differences between the proposed PG--GRU feedforward and traditional feedforward methods.
\end{remark}
\begin{remark}
An important aspect of GRU neural networks (GRUNNs) is guaranteeing stability. The stability problem of GRUNNs has been addressed in \cite{bonassi2021stability}, which provides formal stability conditions for GRUNNs. The developed PG--GRU feedforward controller satisfies the conditions for guaranteeing stability (Assumptions 1 and 2 therein) due to normalizing the input data and initializing the internal state at zero.
\end{remark}
\begin{table*}[htbp]
    \centering
    \caption{Comparison between PG--GRU feedforward and traditional feedforward.}
    \label{tab:pggru_traditional}
    \renewcommand{\arraystretch}{1.2}
    \begin{tabular}{|l|p{3.7cm}|p{10cm}|}
    \hline
    \multicolumn{1}{|c|}{\textbf{Aspect}} & \multicolumn{1}{c|}{\textbf{Traditional Feedforward}} & \multicolumn{1}{c|}{\textbf{PG--GRU Feedforward}} \\
    \hline
    \textbf{Model Basis} 
      & Known physics--based inverse (often linear).
      & Combines a known physics--based inverse with a GRU that learns \emph{residual nonlinearities}. \\
    \hline\textbf{Dynamics Coverage} 
      & Known simplified dynamics.
      & Known simplified dynamics and identifiable residual dynamics. \\
    \hline\textbf{Implementation} 
      & Requires detailed system knowledge for fine--tuning.
      & Requires an existing simplified inverse model or FF controller and training a GRU. \\
    \hline\textbf{Computational Cost} 
      & Matrix-vector multiplication. 
      & Matrix-vector multiplication and GRU inference (efficient for average network size). \\
    \hline\textbf{Interpretability}
      & High -- White--box.
      & Relatively high -- Grey--box.\\
    \hline
    \end{tabular}
\end{table*}
\subsection{Data filter for identification and feedforward control}
\label{section filter}
In real--life systems, output measurements are affected by noise or finite resolution problem when incremental encoders are used.
Especially for the inverse identification adopted for the GRU identification, which is known to be noise--sensitive, this can yield potentially divergent training.
Therefore, we adopt a filter to improve the signal--to--noise ratio. 
Let $F(z)$ denote the discrete‐time filter used to smoothen the output $y(k)$. 
A suitable choice is the Savitzky--Golay filter, which is known to preserve high data moments and given as:
\begin{equation}
    \label{eq:Savitzky_Golay}
    F(z) = \sum_{i=-\frac{m-1}{2}}^{\frac{m-1}{2}}C_i z^i,
\end{equation}
where $m$ is an odd number representing the user--designed moving window size, and $C_i$ are the convolution coefficients computed by applying linear least squares to fit data points in each moving window. 

\begin{figure}
    \centering 
    \includegraphics[width=1.0\linewidth]{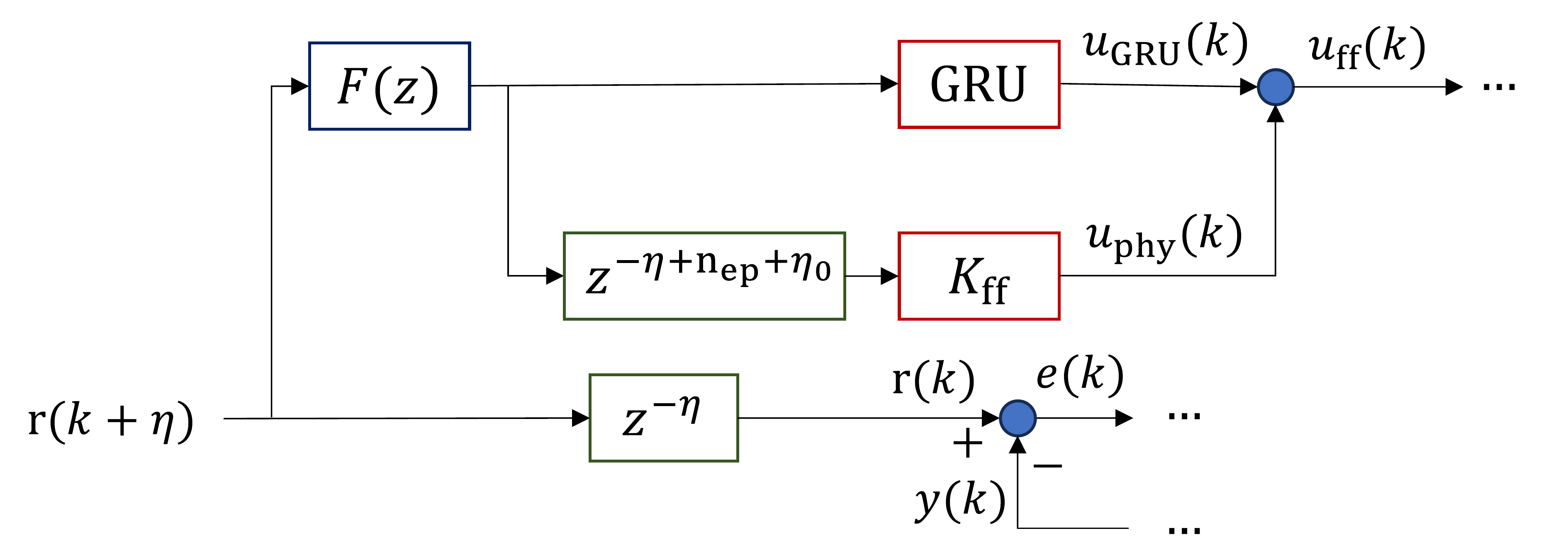}
    \caption{PG--GRU feedforward control design with filter $F(z)$. }
\label{fig:GRU_FF_Implementation}
\end{figure}

In the inverse models~\eqref{eq:InverseModel_Linear} and~\eqref{eq:GRU_Model} we use $F(z) y^d(k)$ instead of $y^d(k)$. 
Moreover, since the identified models now represent $F(z) G^{-1}$, we also need to adjust the feedforward controllers~\eqref{eq:FeedforwardController_TransferFunction},~\eqref{eq:Feedforward_PGGRU} to include the filter, i.e., replace $r(k)$ with $F(z)r(k)$. 
In summary, in~\eqref{eq:InverseModel_Linear},~\eqref{eq:GRU_Model},~\eqref{eq:FeedforwardController_TransferFunction}, and~\eqref{eq:Feedforward_PGGRU} we adjust:
\begin{align}
    \begin{split}
        \label{eq:FilterDesign}
        y^d(k) & \rightarrow F(z) y^d(k),\\
        r(k) & \rightarrow F(z)r(k). 
    \end{split}
\end{align}
The PG--GRU feedforward controller design with filter $F(z)$ is represented in Fig.~\ref{fig:GRU_FF_Implementation}. 
\section{Experimental results}
\label{section results}
\subsection{Two--mass spring--damper system}\label{section Two--mass spring--damper system}
\begin{figure}[t!]
    \centering
    \includegraphics[width=0.85\linewidth]{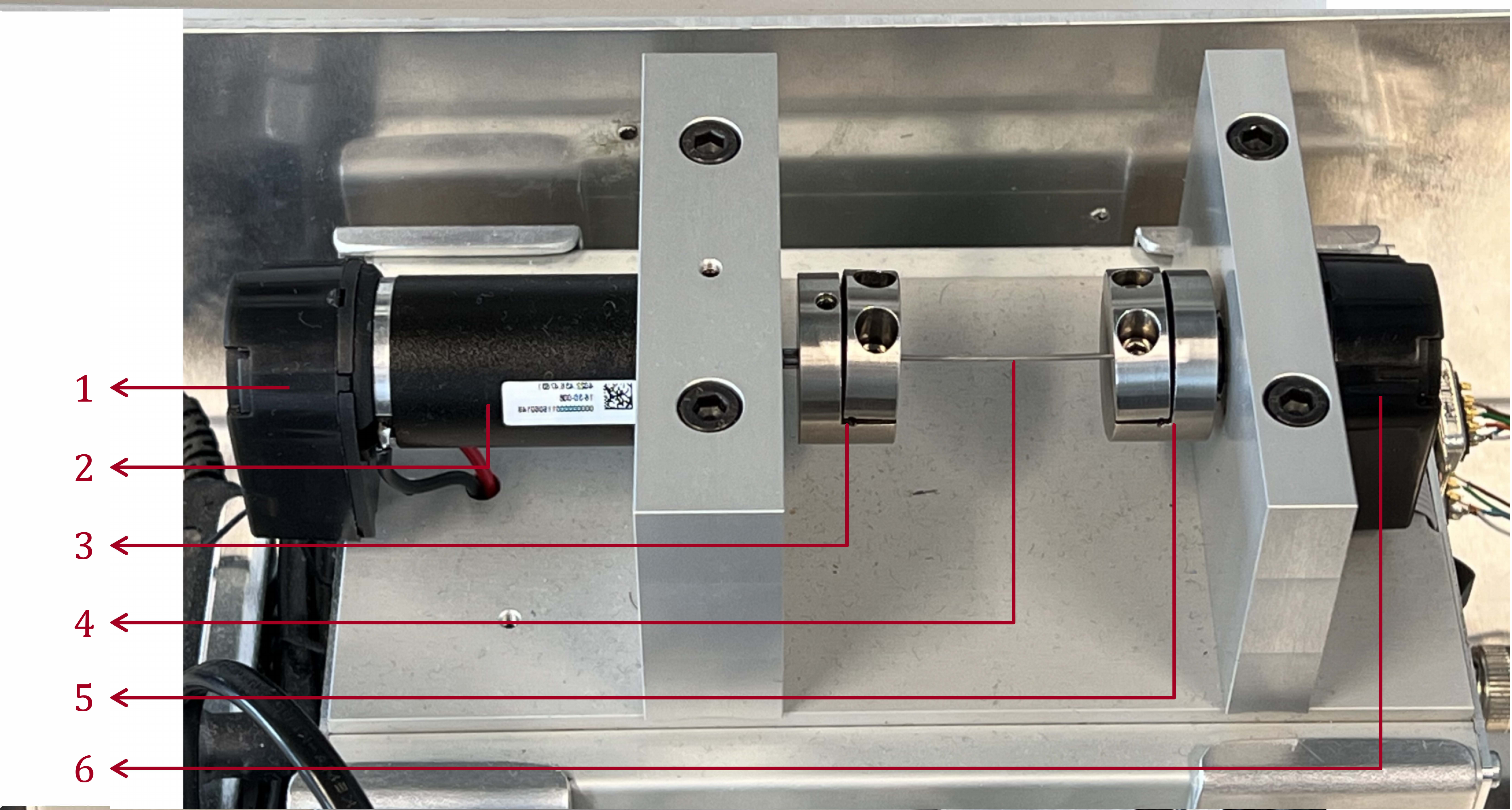}
    \caption{The two--mass spring--damper system: 1. Motor--side encoder; 2. DC motor; 3. Motor inertia; 4. Flexible axle; 5. Load inertia; 6. Load--side encoder.}
    \label{fig setup}
\end{figure}

\begin{figure}[t]
    \centering    
    \includegraphics[width=0.7\linewidth]{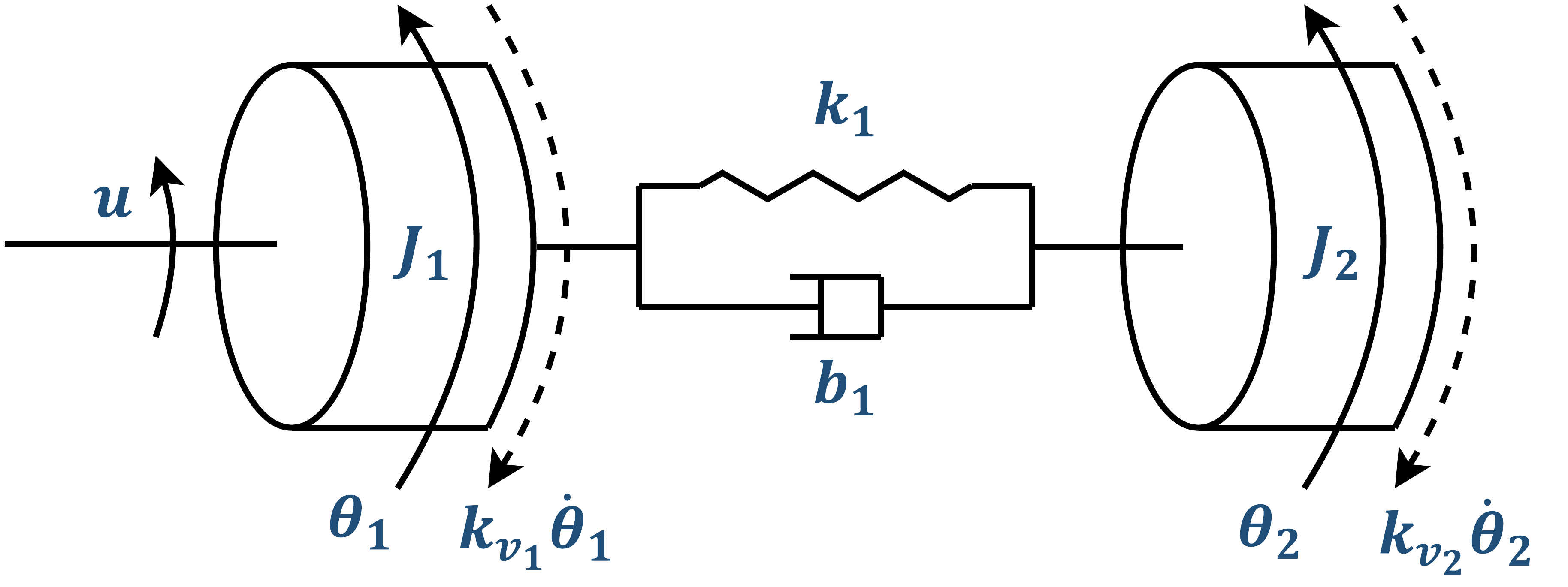}
    \caption{Schematic of the two--mass spring--damper system.}
    \label{fig:Two-Mass Spring Damper}
\end{figure}

We consider the two--mass spring--damper system shown in Fig.~\ref{fig setup} and its simplified version in Fig.~\ref{fig:Two-Mass Spring Damper}. 
The system consists of two rotating masses connected by a flexible axle that is modeled as a spring--damper \cite{kasemsinsup2018experimental}. 
A DC motor applies a torque $u$ to motor inertia, and the objective is to control the rotation of the load inertia. 
Rotations are measured using an encoder with increments of $10^{-3}\pi$ $rad$.
Using Newton--Euler equations yields a linear continuous--time model
\begin{align}
    \begin{split}
        \label{eq:2MSD_Model}
        \dot{x}(t) & = A_c x(t) + B_c u(t), \\
        y(t) & = C x(t), 
    \end{split}
\end{align}
where $t \in \mathbb{R}_{>0}$ denotes the time variable, and
\begin{align}
    \begin{split}
        \label{eq:2MSD_ModelMatrices}
        A_c &= \begin{bmatrix} 0 & 0 & 1 & 0 \\ 0 & 0 & 0 & 1 \\ - \frac{k_1}{J_1} & \frac{k_1}{J_1} & - \frac{b_1 + k_{v_1}}{J_1} & \frac{b_1}{J_1} \\ \frac{k_1}{J_2} & - \frac{k_1}{J_2} & \frac{b_1}{J_2} & - \frac{b_1+k_{v_2}}{J_2} \end{bmatrix}, \quad B_c = \begin{bmatrix} 0 \\ 0 \\ \frac{1}{J_1} \\ 0 \end{bmatrix}, \\
        C &= [0, 1, 0, 0]. 
    \end{split}
\end{align}
In~\eqref{eq:2MSD_Model}, $x(t) = [\theta_1(t), \theta_2(t), \dot{\theta}_1(t), \dot{\theta}_2(t) ]^T$, with $\theta_i(t)$ and $\dot{\theta}_i(t)$ the rotation and angular velocity of mass $i = 1,2$. 
In addition, $k_{v_i}$ and $J_i$ denote the viscous friction coefficient and mass moment of inertia of mass $i=1,2$, respectively, and $b_1$ represents the damping and $k_1$ the stiffness of the flexible axle. 

The two--mass spring--damper system is operated in closed--loop with a sampling time of $T_s = 5\cdot 10^{-4}$~$s$. 
We use zero--order--hold (ZOH) discretization to obtain a discrete--time representation of the continuous--time model~\eqref{eq:2MSD_Model}. 
Moreover, the feedback controller is the Tustin discretization of
\begin{equation}
    \label{eq:FeedbackController}
    K_{\textup{fb}}(s) = 0.007 \left( \frac{ \frac{1}{4 \pi} s + 1 }{\frac{1}{60\pi} s + 1 } \right) \left( \frac{\frac{1}{(90\pi)^2}s^2 + \frac{0.002}{90\pi} s + 1}{\frac{1}{(90 \pi)^2} s^2 + \frac{1}{90\pi} s + 1} \right),
\end{equation}
where $s$ is the Laplace variable. 
The feedback controller is a lead--lag filter in combination with a notch filter. 
\begin{figure}
    \centering 
    \includegraphics[width=1.0\linewidth]{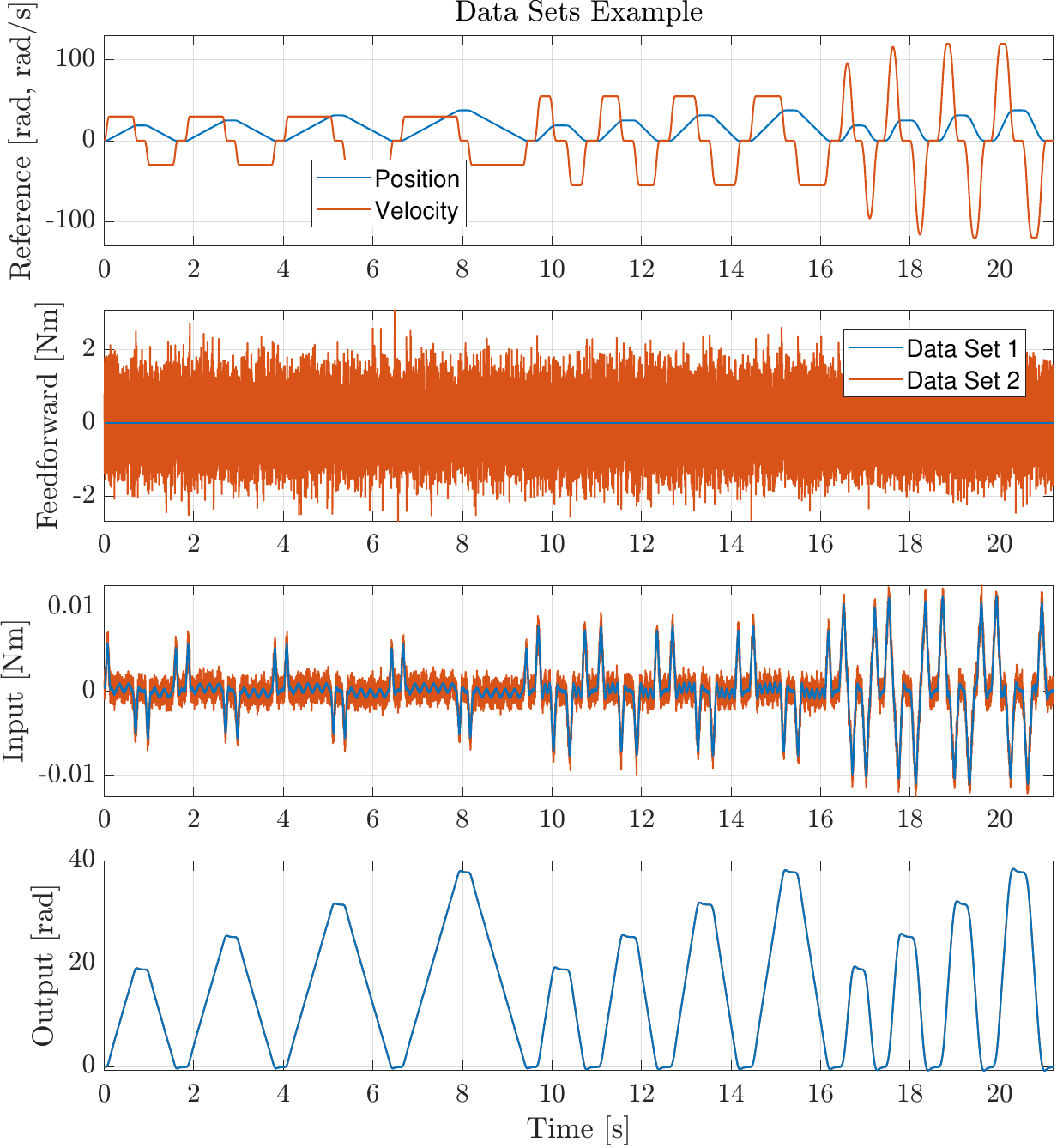}
    \caption{Training data generated on the two--mass spring--damper system.}
\label{fig:Training_DataSets}
\end{figure}

\subsection{Data generating experiment}
\label{subsection data}
We generate the \emph{training data} by operating the two--mass spring--damper system in closed--loop. 
We design the reference trajectory $r^d(k)$ that consists of third--order trajectories moving back and forth from $0$~$rad$ to $6 \pi$, $8\pi$, $10\pi$ and $12\pi$~$rad$. 
These movements are repeated with a maximum velocity of $30$, $55$, and $120$ $\frac{rad}{s}$, and the maximum acceleration is fixed to $1000$~$\frac{rad}{s^2}$. 
Then, reference trajectory $r^d(k)$ is followed twice, with the first repetition using $u_{\textup{ff}}^d(k) = 0$, and the second repetition using a zero--mean white noise with variance $5 \cdot 10^{-7}$~$N^2m^2$ for $u_{\textup{ff}}^d(k)$ to explore a wider range of velocities and accelerations. 
The data sets are visualized in Fig.~\ref{fig:Training_DataSets}. 

Additionally, we generate a \emph{validation data set} separately on the system while using $u_{\textup{ff}}^d(k) = 0$ and a reference trajectory of the following three references sequentially:
\begin{enumerate}
    \item \emph{Slow reference:} from rotation $0$~$rad$ to $6\pi$~$rad$, with velocity $40$~$\frac{rad}{s}$ and acceleration $700$~$\frac{rad}{s^2}$;
    \item \emph{Nominal reference:} from rotation $0$~$rad$ to $7\pi$~$rad$, with velocity $60$~$\frac{rad}{s}$ and acceleration $800$~$\frac{rad}{s^2}$;
    \item \emph{Fast reference:} from rotation $0$~$rad$ to $10\pi$~$rad$, with velocity $100$~$\frac{rad}{s}$ and acceleration $900$~$\frac{rad}{s^2}$. 
\end{enumerate}

\subsection{Feedforward controllers}
We evaluate three feedforward controllers:
\begin{enumerate}
    \item Linear feedforward control with stable inversion;
    \item Preview--based GRU feedforward control;
    \item PG--GRU feedforward control.
\end{enumerate}

The linear feedforward controller~\eqref{eq:FeedforwardController_TransferFunction} is derived from the ZOH discretization of the model~\eqref{eq:2MSD_Model}. 
The parameters are identified according to the feedforward control--oriented identification~\eqref{eq:Identification_Criterion}. 
The relative degree is $\eta_0 = 1$. 
One non--minimum phase zero occurs due to the discretization, which is stable approximated using ZPETC as in~\eqref{eq:StableInversion} which yields $n_{\textup{ep}} = 1$. 

For the GRUs, we adopt a Savitzky--Golay filter $F(z)$ as in~\eqref{eq:FilterDesign} of order $3$ with a window size of $141$ samples to reduce the quantization effects that are induced by the incremental encoder.
The filter is adopted twice.
The GRUs are trained in Pytorch using ADAM optimizer \cite{kingma2014adam}. 
We adopt truncated backpropagation through time (TBPTT) to reduce the problems of vanishing and exploding gradients caused by long data sequences.
A random search is performed for tuning the hyperparameters \cite{bergstra2012random}, where the grid points are highlighted in Table~\ref{table random grid serach}. 
Unlike traditional grid search, random search selects points in the hyperparameter space at random, making it more efficient in high--dimensional settings.
We select, for the preview--based and the PG--GRU the model that achieved the smallest normalized root mean squared simulation error after training. 
The resulting number of layers and neurons, and the preview window $\eta$ are summarized in Table~\ref{tab:Hyperparameters}. 

\begin{table}
\centering
\caption{Grid points for hyperparameter tuning.}
\label{table random grid serach}
\begin{tabular}{|c|c|}
\hline
\textbf{Hyperparameters}           & \textbf{Grid}                                       \\ \hline
Number of layers              & ${[}1,2,3,4,5,6,7{]}$                         \\ \hline
Number of neurons            & ${[}8,16,32,64,128{]}$                        \\ \hline
$\beta = \eta$                       & ${[}2,8,32,48,64,92,128{]} $               \\ \hline
$\lambda$                 & $10^{-5}\times {[}1,2,4,8{]}$    \\ \hline
TBPTT length     & ${[}299,899,1399,2099{]} $                \\ \hline
Learning rate             & $10^{-4}\times {[}1,2,4,8,16{]}$ \\ \hline
Maximum gradients norm & ${[}0.1,0.2,0.4,0.8{]} $                      \\ \hline
Batch size               & ${[}2,4,6{]} $                                \\ \hline
Initialization type       & {[}"Kaiming" \cite{he2015delving}, "Xavier" \cite{glorot2010understanding}{]}                   \\ \hline
\end{tabular}
\end{table}

\begin{table}
    \centering
    \caption{Hyperparameter choices for GRU and PG--GRU.}
    \label{tab:Hyperparameters}
    \begin{tabular}{|c|c|c|} \hline
    & \textbf{GRU} & \textbf{PG--GRU} \\ \hline
    Number of layers & 5 & 7 \\ \hline
    Number of neurons & 128 & 32 \\ \hline
    $\beta = \eta$ & 92 & 48 \\ \hline
    \end{tabular}
\end{table}
\begin{table}[t!]\centering
\caption{NRMS of inverse model identification.}\label{table nrms}
\begin{tabular}{|c|c|c|c|}
\hline
Inverse model& $R_1$   &  $R_2$ &  $R_3$   \\ \hline
Linear& $11.2\%$ & $9.28\%$ & $6.84\%$ \\ \hline
GRU& $27.03\%$ & $19.22\%$ & $17.24\%$ \\ \hline
Preview--based GRU& $10.65\%$ & $9.40\%$ & $7.16\%$ \\ \hline
PG--GRU& $4.49\%$ & $3.75\%$  & $2.68\%$\\ \hline
\end{tabular}
\end{table}
\begin{table}[t!]
\centering
\caption{IAE [rad] of the tracking error resulting from different feedforward controllers.}
\label{tab:RMSE}
\begin{tabular}{|c|c|c|c|}
\hline
$\cdot 10^{-2}$ & $R_1$   &  $R_2$ &  $R_3$   \\ \hline
No feedforward& $22.90$ & $32.44$ & $50.57$ \\ \hline
ZPETC & $2.20$ & $2.82$ & $2.75$ \\ \hline
GRU         & $5.04$            & $4.43$                              & $7.53$      \\ \hline
Preview--based GRU     & $2.23$            & $3.70$                             & $4.39$     \\ \hline
PG--GRU  & $1.28$            & $1.57$                             & $1.93$     \\ \hline
\end{tabular}
\end{table}

\begin{figure*}[h!]
    \centering
    \begin{subfigure}{1\linewidth}
        \centering
        \includegraphics[width=\linewidth]{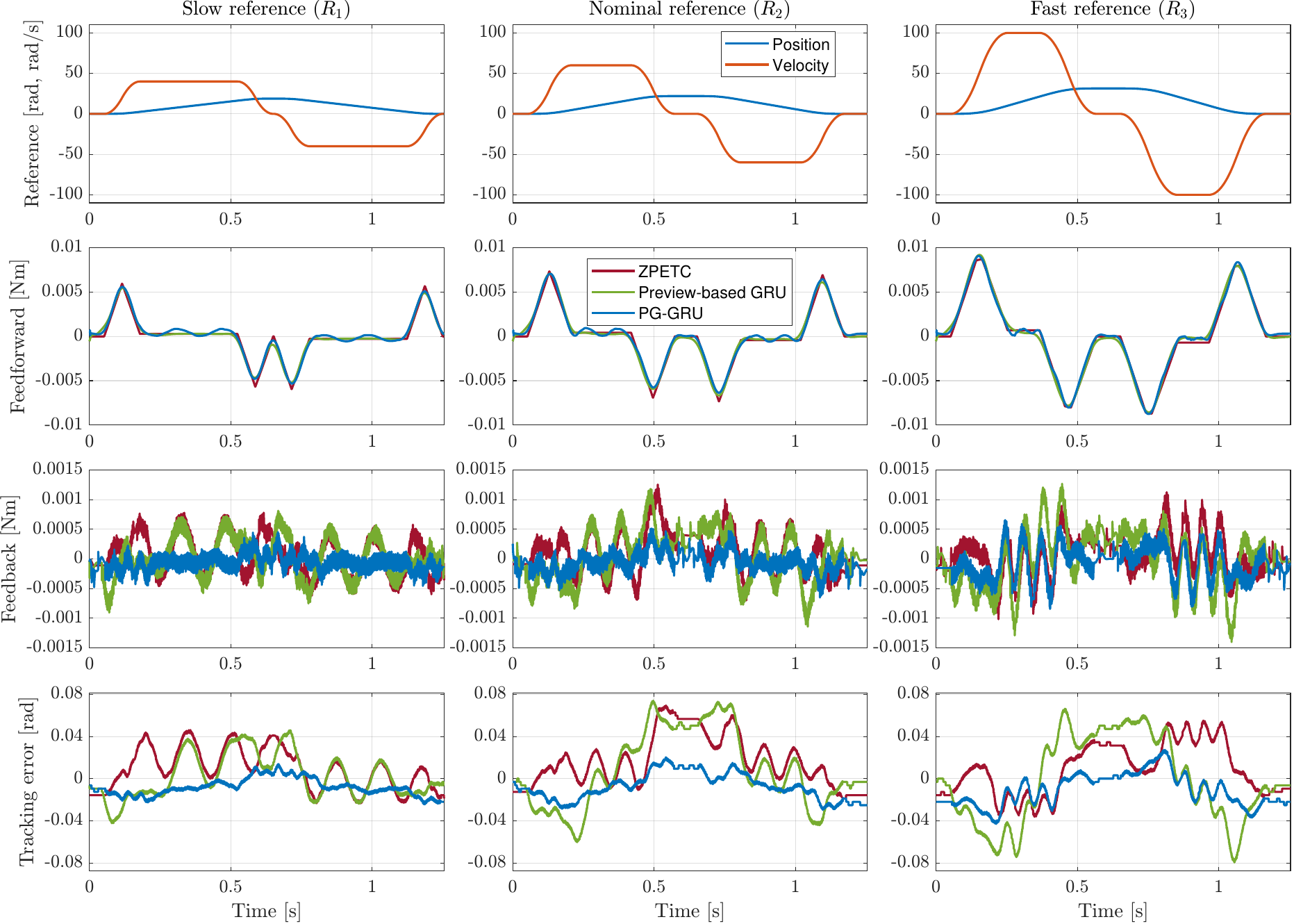}
        \caption{Reference, feedforward input, feedback input and the tracking error resulting from the linear feedforward controller with ZPETC ({\color{Maroon}\textbf{{---}}}), preview--based GRU ({\color{LimeGreen}\textbf{{---}}}), and PG--GRU ({\color{NavyBlue}\textbf{{---}}}) feedforward controller.}
        \label{fig FF all}
    \end{subfigure} 
    \begin{subfigure}{1.0\linewidth}
        \includegraphics[width=\linewidth]{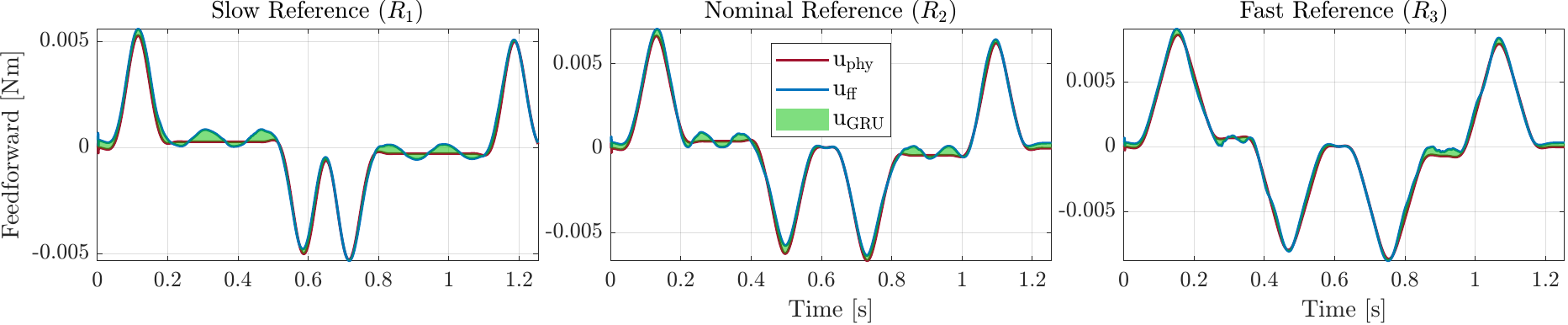}
        \caption{Contributions of the components in the PG--GRU feedforward signal: linear ZPETC feedforward $u_\text{phy}$ ({\color{Maroon}\textbf{{---}}}), the total feedforward input $u_{\textup{ff}}$ ({\color{NavyBlue}\textbf{{---}}}), and GRU contribution $u_\text{GRU}$  ({\color{YellowGreen}\textbf{{$\blacksquare$}}}).}
        \label{fig FF contributions}
    \end{subfigure}
    \caption{Feedforward control results for the slow, nominal and fast reference resulting from linear feedforward with ZPETC, preview--based GRU feedforward and PG--GRU feedforward. }
    \label{fig:Feedforward_Results}
\end{figure*}
\subsection{Feedforward control performance}
Fig.~\ref{fig:Feedforward_Results} shows the control performance on the slow, nominal, and fast reference when using stable inversion, the preview--based GRU, and the PG--GRU feedforward controllers. 
Table~\ref{table nrms} reports the inverse model identification results on validation sets, and Table~\ref{tab:RMSE} summarizes the tracking integral absolute error (IAE) for all references, including also the no--feedforward case and the standard (no--preview) GRU.
Therein, we observe that the lack of preview of the standard GRU makes it perform significantly worse compared to the linear approach. 
In contrast, the preview--based GRU performs closer to linear feedforward controller. 
Most importantly, the PG--GRU significantly outperforms the alternatives and reaches roughly a twofold improvement in IAE on all three references. 
This is caused by the fact that the PG--GRU starts from the linear ZPETC feedforward and uses the GRU only to improve. 
When the velocity reaches its peak, the linear feedforward ($u_{\text{phy}}$) can only compensate for the linear part, while the system dynamics are dominated by velocity--dependent nonlinearities.
At this stage, the GRU component provides effective compensation, greatly reducing the tracking error.
This is visualized in Fig.~\ref{fig FF contributions}, which shows the ZPETC contribution $u_{\textup{phy}}(k)$ and the GRU contribution $u_{\textup{GRU}}(k)$. 
The GRU learns only from the residuals and thereby has a relatively small but critical contribution to handle nonlinearities. 

\section{Conclusions and Discussions}
\label{section conclusions and discussions}
This work developed a PG--GRU architecture for inversion--based feedforward control. Traditional feedforward controllers, such as those based on linear models, often fail to compensate for the non--linearities present in real--world systems. GRUs offer a promising alternative due to their ability to handle nonlinearities and long--term dependencies, but suffer from transients of internal states and lack interpretability due to their black--box nature. 

The proposed PG--GRU framework addressed these limitations by integrating a preview--based GRU with a linear stable inverse model. Experimental validation on a real--life two--mass spring--damper system demonstrated the effectiveness of the PG--GRU feedforward controller. 
Therein, the PG--GRU outperformed both a linear ZPETC feedforward controller and a GRU feedforward controller with preview. 
These results confirmed that the PG--GRU effectively leverages the strengths of both linear and nonlinear modeling techniques, providing superior feedforward performance by compensating for linear and nonlinear dynamics. 
Hence, we conclude that the PG--GRU framework offers a promising direction for future research in high--precision motion control systems with complex state-space nonlinearities. 

PG--GRU converges much faster than a standalone GRU in our experiments, but it still depends on the network’s ability to learn meaningful system residual dynamics from training data. Stability of PG--GRU feedforward controllers is enhanced by incorporating a stable physics--based feedforward controller and it can be analyzed using frameworks of \cite{bonassi2021stability, bolderman2024physics}. The sufficient conditions developed therein are satisfied by the developed PG--GRU FF controllers. Future work will consider the optimal dimensioning of PG--GRUs.
\bibliographystyle{ieeetr}
\bibliography{arxiv.bib}
\end{document}